\documentclass[5p]{elsarticle}

\usepackage{graphicx}% Include figure files
\usepackage{dcolumn}% Align table columns on decimal point
\usepackage{bm}% bold math
\usepackage{hyperref}% add hypertext capabilities
\usepackage{color}
\usepackage{breakurl}
\usepackage{amsmath}
\usepackage{amssymb}

\newcommand{\be}{\begin{equation}}
\newcommand{\ee}{\end{equation}}
\newcommand{\bea}{\begin{eqnarray}}
\newcommand{\eea}{\end{eqnarray}}
\newcommand{\ds}{{\sf DarkSUSY}}

\newcommand{\eref}[1]{Eq.~\eqref{#1}}
\newcommand{\fex}{\textit{e.g.}}

\begin{document}

\title{Indirect dark matter searches as a probe of degenerate particle
spectra}

\author[HH]{Masaki Asano}
\ead{masaki.asano@desy.de}
\author[HH]{Torsten Bringmann}
\ead{torsten.bringmann@desy.de}
\author[MPPMU]{Christoph Weniger}
\ead{weniger@mppmu.mpg.de}

\address[HH]{\mbox{{II.} Institute for Theoretical Physics, University of Hamburg,
Luruper Chaussee 149, DE-22761 Hamburg, Germany}}
\address[MPPMU]{Max-Planck-Institut f\"ur Physik, F\"ohringer Ring 6, 80805 Munich, Germany}

\begin{abstract}
  We consider the possibility that the cosmological dark matter consists of
  particles very close in mass to new colored particles below the TeV scale.
  While such a scenario is inherently difficult to directly confirm at
  colliders, we find that indirect dark matter searches may be a powerful
  alternative. In particular, we show that in this case dark matter
  annihilation to $\bar q q g$ final states can give rise to significant
  antiproton (but also gamma-ray) fluxes, and compare the resulting
  constraints to bounds from direct searches at LEP, the Tevatron and the LHC.
  For supersymmetric neutralinos degenerate with squarks, \fex, antiprotons
  can give rise to more stringent constraints for masses below around
  45--80~GeV.
\end{abstract}

\maketitle

%%%%%%%%%%%%%%%%%%%%%%%%%%%%%%%%%%%%%%%%
%%%%%%%%%%%%%%%%%%%%%%%%%%%%%%%%%%%%%%%%
\section{Introduction}
Collider experiments have pushed the scale for possible new physics beyond the
standard model (BSM) to ever higher energies in recent years. The CERN LHC, in
particular, is now running at a center of mass energy of $7\,$TeV and the
non-observation of any clear BSM signal, so far, has allowed to place
impressive limits~\cite{Aad:2011qa,Chatrchyan:2011zy}; after an integrated
luminosity of $1.04$ fb$^{-1}$, scalar quarks (squarks) in  supersymmetric
extensions of the standard model, \fex, have already been excluded for masses
below $\sim 900$ GeV~\cite{Aad:2011ib,cmssimp}.  One has to keep in mind,
however, that these results rest on the assumption that the new colored states
quickly decay into the lightest neutralino, which is assumed to be massless
for the sake of the analysis, and emit high-energy QCD jets in this process.
These limits thus do not apply if the only accessible lighter states are very
close in
mass~\cite{ATLAS:limit,Chatrchyan:2011ek,LeCompte:2011cn,LeCompte:2011fh}, in
which case the resulting jets and missing transverse energy would be too soft
to pass the signal selection criteria.  Scenarios with degenerate particle
spectra are thus generically very difficult to probe at hadron colliders; the
situation is considerably better for electron-positron colliders like LEP, but
even in this case the limits from direct searches for colored states do not
apply to highly degenerate spectra~\cite{Heister:2002hp, Abbiendi:2002mp,
Achard:2003ge, Abdallah:2003xe, LEP_SUSYWG}.

Here, we investigate whether such a situation could be probed by indirect dark
matter (DM) searches, assuming that the lightest BSM state is a weakly
interacting massive particle (WIMP) that makes up the cosmological DM.  The
self-annihilation of WIMP pairs  could then leave an imprint in the spectrum
of cosmic rays (see, \fex, Ref.~\cite{Bertone:2004pz} for a review on particle
DM and indirect searches).  As we will demonstrate, WIMP annihilation into
$\bar qq g$ final states, a channel so far hardly explored in indirect DM
searches, may give rise to the most stringent constraints in degenerate
scenarios (mostly through antiproton, but also through gamma-ray production).
In fact, this channel may help to fill remaining loop-holes for the existence
of new colored states below masses of around 100 GeV that are left from direct
collider searches for such particles. 

The rest of this Letter is organized as follows: In Section 2 we discuss the
importance and main characteristics of DM annihilation into $\bar{q}qg$ final
states, focussing on the case of neutralino DM; in Section 3 we present the
energy spectra of antiprotons that result from $\bar{q}qg$ hadronization and
discuss antiproton propagation through our Galaxy; Section 4 contains the
limits on the annihilation cross-section that can be derived from cosmic-ray
observations and Section 5 confronts these limits with direct collider
searches for charged colored particles. Finally, in Section 6 and 7,
respectively, we present a discussion of our results and conclude.

%%%%%%%%%%%%%%%%%%%%%%%%%%%%%%%%%%%%%%%%
%%%%%%%%%%%%%%%%%%%%%%%%%%%%%%%%%%%%%%%%
\section{Dark matter annihilation into gluons}

Sizable  cosmic-ray fluxes from galactic DM can only be expected from $s$-wave
annihilation processes; this is  because the $p$-wave rate scales like $v^2$
and is  heavily suppressed for the typically expected DM velocities of the
order of $v\sim10^{-3}$. However, if the annihilating DM pair is in a $J=0$
state---as is necessarily the case for scalar or Majorana DM particles
$\chi$---the tree-level $s$-wave annihilation rate into fermion pairs $\bar
ff$ is helicity suppressed and scales like $\sigma v\propto m_f^2/m_\chi^2$.
In this situation, the existence of an additional vector boson in the final
state may lift the helicity suppression and lead to a radiative 'correction'
of the order of $(\alpha/\pi)\,m_\chi^2/m_f^2$, which in practice can
correspond to an annihilation rate that is enhanced by several orders of
magnitude~\cite{Bergstrom:1989jr,Flores:1989ru}.

\begin{figure}[t]
  \includegraphics[width=\columnwidth]{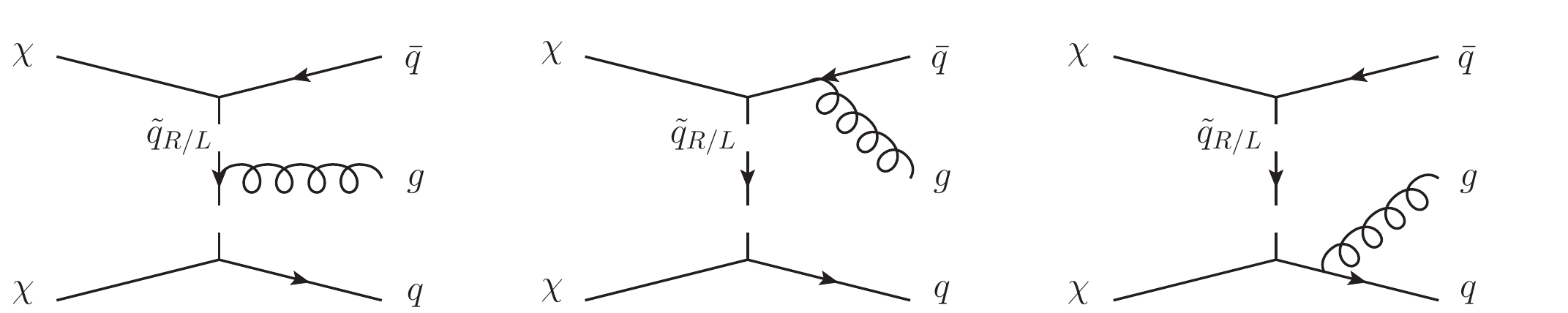}
  \caption{Feynman diagrams contributing to $\chi\chi\rightarrow \bar q qg$ in
  the limit of vanishing quark mass and neutralino velocity.}
  \label{fig:feynman} 
\end{figure}

We will in the following focus on the supersymmetric neutralino, but note that
we expect qualitatively very similar results for other DM candidates where the
annihilation rate into fermions is helicity suppressed.  The annihilation of
neutralinos into $\bar q q$ two-body final states, in particular, is typically
strongly suppressed (at least for $m_b\ll m_\chi<m_t$ or $m_\chi\gg m_t$), so that
the dominant contribution to the total annihilation cross section into quarks
is given by the process $\chi\chi\rightarrow \bar q qg$. In the
$m_q\rightarrow0$ limit, only the squark-exchange $t$-channel diagrams shown
in Fig.~\ref{fig:feynman} contribute and we find
\begin{align}
  \label{sigtot}
  &(\sigma v)^{\chi\chi\rightarrow \bar q qg}_{v\rightarrow0} =
  \frac{\alpha_\mathrm{s}\left|\tilde
  g_R\right|^4}{16\pi^2m_\chi^2}\times\Bigg\{\frac{3\!+\!4{\mu_R}}{1\!+\!{\mu_R}}\nonumber\\
  &\quad+\left(1\!+\!{\mu_R}\right)\!\left[\frac{\pi^2}{6}-\!
  \left(\!\log\frac{{\mu_R}\!+\!1}{2{\mu_R}}\right)^2\!-2\mathrm{Li}_2\!\left(\frac{{\mu_R}\!+\!1}{2{\mu_R}}
  \right)\right]\nonumber\\
  &\quad+\frac{4{\mu_R}^2\!-\!3{\mu_R}\!-\!1}{2{\mu_R}}\log\frac{{\mu_R}\!-\!1}{{\mu_R}\!+\!1}\Bigg\}
  +\Big(R\leftrightarrow L\Big)\,,
\end{align}
where   ${\mu_{R,L}}\equiv m_{{\tilde q}_{R,L}}^2/m_\chi^2$ and $\tilde
g_RP_L$ ($\tilde g_LP_R$) is the coupling between neutralino, quark and
right-handed (left-handed) squark;  $\mathrm{Li}_2(z)=\sum_{k=1}^\infty
z^k/k^2$  is the dilogarithm. This result is consistent with the differential
cross section presented in Ref.~\cite{hep-ph/9306325}; for comparison, it can
also be obtained by the corresponding expressions for photon final states
$\bar qq\gamma$~\cite{arXiv:0808.3725} by simply replacing $Q^2\alpha_{\rm
em}\rightarrow (4/3)\,\alpha_s$, where $Q$ is the electric charge of
$q$.\footnote{Note that there is a typo in Eq.~(2) of
Ref.~\cite{arXiv:0808.3725}---see also Ref.~\cite{Bell:2011if}.}

If the neutralino is a pure Bino with $m_b\ll m_{\tilde B}<m_t$, the
\emph{total} annihilation cross section is well approximated by \eref{sigtot}
as long as the mass-splitting $\mu$ remains small. This is thus the
\emph{minimal} neutralino annihilation cross section we can expect; in the
exactly degenerate case, $\mu_R=\mu_L=1$, it is largest and becomes
\be
  (\sigma v)^{\tilde B\tilde B\rightarrow \bar q qg}_{v\rightarrow0} =
  \frac{2\alpha_{\rm s}\alpha^2_Y}{3m_{\tilde B}^2}\left(21-2\pi^2\right)\left(6^{-4}+Q^4\right)\,.
\ee
For other neutralino compositions, \fex~if the neutralino is a Wino, one can
find higher total cross sections---mainly because annihilation into $W^+W^-$
and $ZZ$ final states becomes effective (but also because the
neutralino-(s)quark couplings in \eref{sigtot} can be larger). However, these
scenarios tend to be more constrained than Bino neutralinos, and we will here
concentrate on the latter.

%%%%%%%%%%%%%%%%%%%%%%%%%%%%%%%%%%%%%%%%
%%%%%%%%%%%%%%%%%%%%%%%%%%%%%%%%%%%%%%%%
\section{Antiproton production and propagation}

\begin{figure}[t]
  \includegraphics[width=\columnwidth]{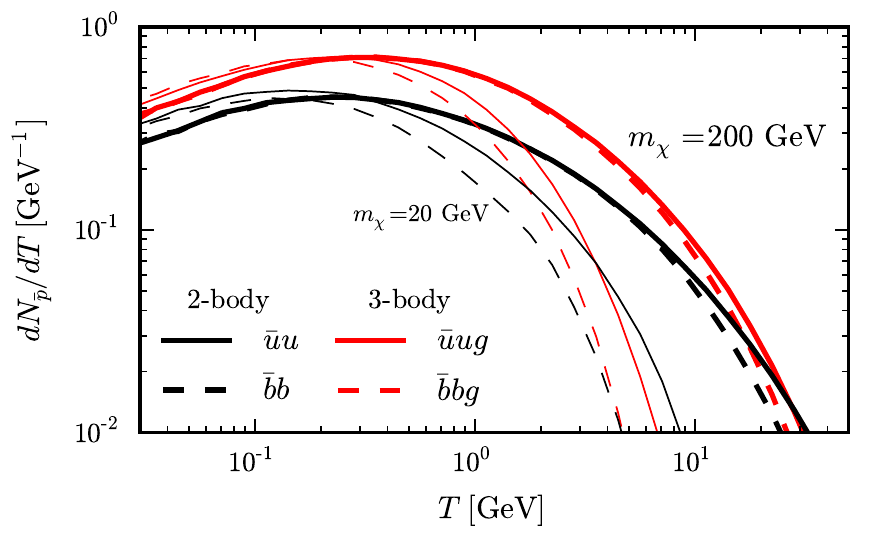}
  \caption{Differential number of antiprotons, \emph{per annihilation},  from
  various channels and as  function of the antiproton kinetic energy $T$. Red
  (black) curves show the result from three-body (two-body) final states
  containing up-quarks (solid lines) and bottom-quarks (dashed lines),
  respectively. For thick lines, the DM mass is set to 200\,GeV, for thin
  lines to 20\,GeV.  We adopted a mass splitting of $\mu=1.2$, but the spectra
  are practically independent of $\mu$ in the range $\mu=1\dots4$.}
  \label{fig:dNdx} 
\end{figure}

The fragmentation of quarks and gluons into color singlets leads to the
production of a sizable amount of antiprotons. We derived the antiproton
energy spectrum $dN_{\bar{p}}/dT$ (with $T$ denoting kinetic energy) from DM
annihilation using Monte Carlo methods.  To this end, we simulated the
distribution of hard partons from $\chi\chi\to \bar q q g$, following our
above analytical results, and used the event generator \textsf{Pythia
6.4.19}~\cite{hep-ph/0603175} to perform the subsequent parton showering,
fragmentation and particle decay.

Our results for the antiproton energy spectra are shown in Fig.~\ref{fig:dNdx}
for different DM annihilation channels and $\mu=1.2$ (assuming that $\mu
\equiv \mu_L \equiv \mu_R$); for other values of $\mu$ we obtain very similar
results. We find that, in the phenomenologically relevant low energy region,
the additional hard gluon in $\bar{q}qg$ final states leads to an enhancement
of antiproton production by a factor of up to $\sim2$ relative to the
$\bar{q}q$ final states commonly considered in  indirect DM searches. Note
that while final state \emph{electroweak} gauge bosons at first sight can lead
to a much larger enhancement of $dN_{\bar p}/dT$ for low-energy
antiprotons~\cite{arXiv:1009.0224}, this  is only because  final state
radiation of gluons is already included in the commonly adopted 'two-body'
result obtained from \textsf{Pythia}---while that of electroweak gauge bosons
is not. Fig.~\ref{fig:dNdx} also shows that the amount of antiprotons
increases with decreasing quark mass, as is expected for final state radiation
(which is dominated by collinear gluons).

Once produced, antiprotons do not travel along straight lines  like, \fex,
gamma rays, but  scatter on randomly distributed galactic  magnetic field
inhomogeneities. Their propagation  can thus nicely be described in terms of a
phenomenological diffusion model, the free parameters of which  are  strongly
constrained by other cosmic ray data, in particular the boron over carbon
ratio $B/C$~\cite{astro-ph/0306207}. This allows to predict the expected
astrophysical background, consisting of \emph{secondary} antiprotons mainly
produced in the collisions of cosmic ray protons with the interstellar medium,
with a remarkable accuracy~\cite{Donato:2001ms,astro-ph/0612514}.  The recent
cosmic-ray antiproton flux measurements undertaken by the PAMELA experiment
over rigidities of 0.35--180 GV~\cite{arXiv:1007.0821} are fully consistent
with this background.

The expected flux of \emph{primary} antiprotons from DM
annihilation~\cite{NSF-ITP-84-80}, on the other hand, is subject to  greater
uncertainties~\cite{astro-ph/0306207}. The reason for this is that the $B/C$
analysis, being restricted to sources in the galactic disk, actually leaves
large degeneracies in the diffusion parameters---in particular between the
size of the diffusion constant and the thickness $L$ of the diffusion zone
perpendicular to the galactic plane. While secondary antiprotons are not
greatly affected by this degeneracy, DM annihilation happens predominantly in
the galactic halo; therefore,  a much larger volume of the diffusion zone is
probed and the antiproton flux is rather sensitive to the adopted value of
$L$.  Even though the $B/C$ analysis in principle allows values as small as
$L\sim1\,$kpc, values up to $L\sim10\,$kpc are preferred when adding
radioactive isotopes to the analysis~\cite{arXiv:1001.0551}, with similar
conclusions following from the consideration of gamma
rays~\cite{arXiv:1102.0744}  and cosmic-ray electrons~\cite{electronL}. Also
radio data are in rather strong conflict with a halo size as small as
$\sim1\,$kpc~\cite{arXiv:1106.4821}. In our analysis, we will therefore refer
to the 'KRA' model of Ref.~\cite{Evoli:2011id} and the $B/C$ best fit ('MED')
model of Ref.~\cite{astro-ph/0306207}, both featuring $L=4\,$kpc, as well as
the 'MAX' model of Ref.~\cite{astro-ph/0306207}, featuring $L=10\,$kpc.

%%%%%%%%%%%%%%%%%%%%%%%%%%%%%%%%%%%%%%%%
%%%%%%%%%%%%%%%%%%%%%%%%%%%%%%%%%%%%%%%%
\section{Indirect detection constraints for $\bar q qg$ final states}

\begin{figure}[t]
  \includegraphics[width=\columnwidth]{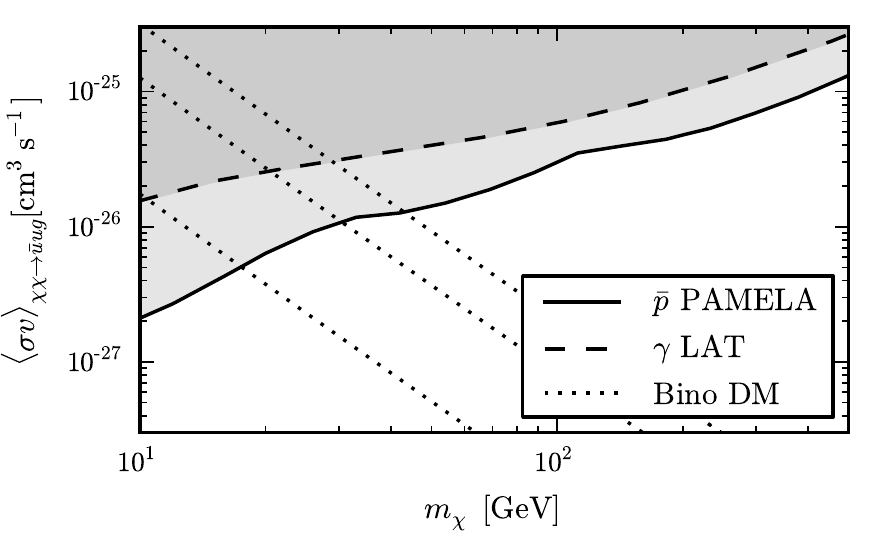}
  \caption{Limits on the DM annihilation cross-section $\langle\sigma
  v\rangle$, assuming a 100\% branching ratio into $\bar{u}ug$ final states.
  Gamma-ray limits (\textit{dashed line}) are taken from  Fermi LAT dwarf
  galaxy observations~\cite{Ackermann:2011wa} and antiproton limits
  (\textit{solid line}) are derived from PAMELA data~\cite{arXiv:1007.0821}.
  \textit{Dotted lines} show the theoretically expected values for Bino DM and
  mass splittings of $\mu\equiv m_{\tilde u}^2/m_{\tilde B}^2=1.01, 1.2,2.0$
  (from top to bottom); the cosmic-ray limits are practically independent of
  $\mu$.}
  \label{fig:svlim} 
\end{figure}

We calculated the antiproton flux from DM annihilation, as expected at the top
of the atmosphere after propagation, by using \ds~\cite{DS} (which implements
the procedure described in Refs.~\cite{Evoli:2011id, Bergstrom:1999jc}).
Solar modulation is taken into account by adopting the force-field
approximation, with a Fisk potential of $\phi=500\,$MV, but our limits
actually do not strongly depend on this choice because they typically do not
derive from antiproton energies much smaller then 1\,GeV. We choose  an
Einasto profile to model the galactic DM halo, with parameters as used in
Ref.~\cite{Evoli:2011id}; for a Navarro-Frenk-White profile, we get very
similar results. Furthermore, we assume that the neutralino makes up all of
the observed dark matter (noting that, due to the potentially very efficient
coannihilation between squarks and neutralinos, this can require a non-thermal
production mechanism or a non-standard freeze-out history). 

In order to derive limits on the flux of primary antiprotons from DM
annihilation we require that the DM signal plus the background of secondary
antiprotons does not overshoot the antiproton measurements by
PAMELA~\cite{arXiv:1007.0821} by more than $3\sigma$ in any data point; to be
conservative, we adopt the \emph{minimal} astrophysical background flux from
Ref.~\cite{astro-ph/0612514}.  The resulting limits on the annihilation
cross-section $\langle\sigma v\rangle$ are shown in Fig.~\ref{fig:svlim}, as
function of the DM mass $m_\chi$, for the case of annihilation into
$\bar{u}ug$ final states and adopting a mass splitting of $\mu=1.2$; we obtain
very similar results for other values of $\mu=1\dots4$ and for the light quark
flavors $q=d,s$, but the limits are up to a factor of two weaker in case of
the heavier quarks $q=c,b$. We note that these limits are in rather good
agreement with previous results, like in Ref.~\cite{Evoli:2011id}, for
two-body final states.

Besides antiprotons, the hadronization of quarks and gluons also generates  a
large amount of gamma rays, which mostly stem from $\pi^0$-decay.  Using again
\textsf{Pythia}, we find reasonable agreement between the gamma-ray energy
spectrum from $\bar{q}qg$ and the commonly adopted $\bar{b}b$ spectrum (at the
peak of the spectra at $E_\gamma\approx m_\chi/20$ the fluxes agree within
$20\%$ for $q=d,u,s,c,b$). Currently, the Fermi Large Area Telescope (LAT) is
measuring the gamma-ray sky with unprecedented precision and strong limits on
the DM annihilation cross section were derived from observations of dwarf
galaxies~\cite{Abdo:2010ex,GeringerSameth:2011iw,Ackermann:2011wa}, galaxy
clusters~\cite{Ackermann:2010rg,Huang:2011xr} and the isotropic diffuse
gamma-ray backround~\cite{Abdo:2010dk}. In Fig.~\ref{fig:svlim} we show the
corresponding limits on the annihilation cross-section to $\bar{q}qg$ as
derived from the Fermi LAT data by a combined analysis of ten dwarf spheroidal
galaxies~\cite{Ackermann:2011wa}, making use of  the above-mentioned
similarity between the gamma-ray spectra from $\bar q qg$ and $\bar b b$ final
states.  It is reassuring that these limits are almost as strong as our
antiproton bounds, since limits derived from antiproton and gamma-ray
observations are in principle subject to very different astrophysical
uncertainties.

For comparison, we also indicate in Fig.~\ref{fig:svlim} the value of the
annihilation cross-section for $\chi\chi\to \bar{u}ug$ in the case of pure
Bino DM, as function of the neutralino mass $m_\chi$ and the mass splittings
$\mu$. Depending on the value of $\mu$, we find that DM masses up to $\sim50$
GeV can be excluded by antiproton and gamma-ray limits; this will be
confronted with collider limits in the next section.

%%%%%%%%%%%%%%%%%%%%%%%%%%%%%%%%%%%%%%%%
%%%%%%%%%%%%%%%%%%%%%%%%%%%%%%%%%%%%%%%%
\section{Collider constraints on new colored particles}

\begin{figure}[t!]
  \includegraphics[width=\columnwidth]{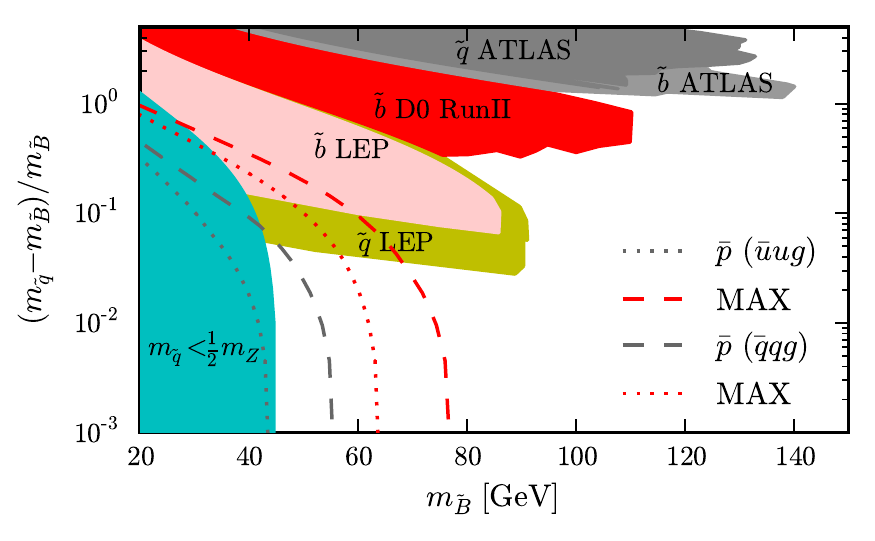}
  \caption{Constraints on squarks degenerate with a Bino LSP, as function of
  the Bino mass and the mass splitting. The colored areas are excluded by
  different collider searches as described in the text and the areas below the
  black lines  are constrained by cosmic-ray antiproton observations (for the
  'MED' propagation model). Dotted lines correspond to light right-handed
  up-type squarks, dashed lines assume degeneracy between all $d,u,c$ and
  $s$-type squarks. The red lines show how the limits strengthen in case of
  the 'MAX' propagation model.}
  \label{fig:muconstraints} 
\end{figure}

In general it is difficult to directly constrain the DM mass with collider
experiments.  On the other hand, light charged and colored particles can be
searched for with many experiments, and in the recent past the LHC has put
strong lower limits on squark masses, reaching up to $\sim 900$
GeV~\cite{Aad:2011ib}. However, these constraints are highly model dependent
and, in the case of supersymmetry,  depend crucially on the supersymmetric
particle mass spectrum: very heavy gluinos, \fex , would suppress the squark
production in hadron colliders.  Additionally, in the parameter region in
which the mass difference between squarks and the lightest superpartner (which
is the DM candidate) is very small, the constraints from collider experiments
are evaded since the energy of QCD jets and missing transverse momentum that
stem from $\tilde{q}\to q\chi$ are small, and these events therefore cannot
pass triggers and cuts.  In general, for light squarks with masses less than
$\sim 100$ GeV, the constrains are very severe because such light particles
are produced at LEP experiments.  Even if the mass splitting is relatively
small, $\sim 5$-$10$ GeV, a large parameter region of the squark mass in which
it can be pair produced is constrained~\cite{Heister:2002hp, Abbiendi:2002mp,
Achard:2003ge, Abdallah:2003xe, LEP_SUSYWG}.  For squarks that are lighter
than $\sim m_Z/2$, the $Z$ boson decay width generally gives the strongest
constraints~\cite{Nakamura:2010zzi}.

Light squarks are also \emph{indirectly} constrained by experiments due to the
contribution of loop diagrams; contributions of these particles have \fex~a
large effect on electroweak precision measurements and flavor changing neutral
currents~\cite{RamseyMusolf:2006vr}.  However, since these limits on the
squark mass are generically very model-dependent,
% and do not directly affect our results, 
we will not discuss them in this Letter.
 
In Fig.~\ref{fig:muconstraints} we show the constraints on squarks degenerate
with a Bino LSP that derive from collider null searches, compared with the
constraints from cosmic-ray antiproton observations. In particular, we show
the following limits: squark and sbottom mass constraints from the LEP
experiments~\cite{Achard:2003ge,LEP_SUSYWG}, sbottom mass constraints from the
Tevatron D0 experiment~\cite{Abazov:2010wq} and constraints from the LHC ATLAS
experiment on squark and sbottom
masses~\cite{ATLAS:limit,Collaboration:2011cw} (for constraints from the CMS
experiment, see also~\cite{cms_squark}).  Concerning the antiproton
constraints, we consider two limiting cases: (1) only the right-handed up-type
squark is light (dotted lines), (2) $d,u,c$ and $s$-type squarks are mass
degenerate and light (dashed lines).  The areas below the corresponding curves
are excluded by antiproton measurements with PAMELA, where we adopted the
'MED' propagation model (for the black curves). 

As expected, these limits are strongest for \textit{small} mass splittings and
are hence constraining the parameter space from a direction that is
complementary to the collider limits. As can be seen from
Fig.~\ref{fig:muconstraints}, in case of light up-type squarks and small
mass-splitting, we can exclude Bino masses up to 45 GeV (the 'KRA' propagation
model gives limits that are about 10 GeV stronger); under favorable
astrophysical conditions (the 'MAX' model) these limits could strengthen to
above 60 GeV. If all $d,u,c$ and $s$-type squarks are mass degenerate, the
lower limits on the Bino mass increase  to around 55--80 GeV, depending on the
astrophysical scenario.  Hence, we find that cosmic-ray observations, in
particular the observations of antiprotons, can be a powerful probe of a
parameter region of DM models that is generically difficult to access directly
with colliders.

%%%%%%%%%%%%%%%%%%%%%%%%%%%%%%%%%%%%%%%%
%%%%%%%%%%%%%%%%%%%%%%%%%%%%%%%%%%%%%%%%
\section{Discussion}

While we have demonstrated that favorable astrophysical conditions, in
particular a large diffusive halo like in the 'MAX' model, could significantly
improve the constraining power of indirect DM searches with antiprotons, we
would like to stress that prospects could be even better when taking into
account the possibility that the DM distribution does not follow a smooth
Einasto profile. For an adiabatically compressed profile as in
Refs.~\cite{Gnedin:2003rj, Gustafsson:2006gr}, which has been argued to result
from the gravitational impact of the observed galactic distribution of
\emph{baryons}  on the DM profile, \fex, we find a further enhancement of the
antiproton flux by a factor of about 4--5 at 1 GeV, which corresponds to an
improvement for the lower bounds on the DM mass by about 20--30 GeV.  Another
rather large enhancement could result from the fact that DM is not expected to
be distributed smoothly, but to cluster in the form of small
subhalos~\cite{Bergstrom:1998jj}.  The resulting effective 'boost-factor' of
the antiproton flux could in principle be quite
large~\cite{astro-ph/0612514,Brun:2007tn,Regis:2009qt}, though one should note
that statistically a much smaller value (corresponding to a factor of a few
enhancement) is expected when extrapolating the results from $N$-body
simulations of gravitational clustering~\cite{Pieri:2009je}.  Here, we will
not study such effects in more detail but simply note that already a moderate
boost-factor of 5 (10) would improve the lower limit on $m_\chi$ by roughly 25
(40) GeV.

Future prospects for indirect DM searches with antiprotons may be even more
promising, given that new data with excellent statistics can soon be expected
from the recently launched AMS-02 experiment~\cite{ams}.  This can only lead
to better limits on exotic contributions to the cosmic-ray antiproton flux,
however, if complemented with an improved understanding of the astrophysical
background. It was estimated that this could strengthen current bounds by
about one order of magnitude~\cite{Evoli:2011id}, which would make it possible
to probe Bino masses up to 100 GeV and beyond, depending on the size of the
mass-split\-ting, the astrophysical propagation model and the nature of the
light squark flavors. In addition, as can be seen from Fig.~\ref{fig:svlim},
gamma-ray observations are also quite constraining for the  channels
considered here; the current limits from Fermi LAT observations will
strengthen further with a better understanding of the Galactic center region
and more accumulated data.

In this Letter we have concentrated on Bino-like neutralinos, which are
probably the most difficult to probe due to their relatively weak interactions
with fermions and their suppressed annihilation signal in cosmic rays.  Let us
stress that  the discussed channel $\bar{q}qg$ can also be relevant for Wino-
or Higgsino-like neutralinos, in which case it is however often dominated by
other two-body channels like, in particular, $\chi\chi\to ZZ$ and $W^+W^-$. In
these cases, masses below a few hundred GeV are difficult to reconcile with
existing gamma-ray or antiproton limits, see \fex~Refs.~\cite{Evoli:2011id,
Ackermann:2011wa}, and we leave a phenomenological exploration of $\bar{q}qg$
final states at $\mathcal{O}(100$ GeV--1 TeV$)$ masses for future work.

Finally, let us  mention another important possibility to search for
degenerate mass spectra. The squark-neutralino degeneracy is also expected to
yield strong constraints from direct dark matter searches because the
neutralino-nucleus scattering cross section can be resonantly enhanced due to
the $s$-channel squark exchange diagram. In fact, the XENON100 experiment can
have sufficient sensitivity for DM masses around $50$
GeV~\cite{Aprile:2011hi}. A detailed analysis of direct detection constraints
for any possible squark-neutralino degeneracy, however, is beyond the scope of
this letter and left for future work (but see Ref.~\cite{arXiv:1110.3719}).

%%%%%%%%%%%%%%%%%%%%%%%%%%%%%%%%%%%%%%%%
%%%%%%%%%%%%%%%%%%%%%%%%%%%%%%%%%%%%%%%%
\section{Conclusions}
In this Letter, we have described a DM annihilation channel, and discussed its
phenomenological implications in some detail, that so far has hardly been
explored in the context of indirect dark matter searches.  If DM is a Majorana
fermion $\chi$, or a scalar, direct annihilation via $\chi\chi\to\bar{q}q$ is
strongly helicity suppressed. In this case, the overall annihilation rate
\emph{today} can be dominated by internal gluon Bremsstrahlung,
$\chi\chi\to\bar{q}qg$, and can be boosted by several orders of magnitude with
respect to the tree-level result;  while this enhancement is much smaller in
the early universe, during \emph{freeze-out}, it can also have a
non-negligible effect on the relic
density~\cite{Flores:1989ru,hep-ph/9306325,hep-ph/0608215}. The precise
branching ratios depend on the difference between the DM particle's mass
$m_\chi$ and the squark mass $m_{\tilde{q}}$; the process
$\chi\chi\to\bar{q}qg$ becomes strongest in the mass degenerate case, i.e.~for
$m_{\tilde{q}}/m_\chi\simeq1$.

We studied the spectra of cosmic rays generated by this  channel, and found
that the energy spectrum of antiprotons coming from $\bar{q}qg$ is generically
enhanced  with respect to the spectrum from $\bar{q}q$ final states
(Fig.~\ref{fig:dNdx}). The corresponding limits on the annihilation
cross-section for $\chi\chi\to\bar{q}qg$ are hence somewhat stronger than
limits on the traditional $\bar{b}b$ or $W^+W^-$ final states
(Fig.~\ref{fig:svlim}). From this figure we can see, as noted
before~\cite{Evoli:2011id,Bottino:2005xy,Bringmann:2009ca,Lavalle:2010yw},
that antiprotons are most constraining for light DM masses, which in
particular seems to disfavor the simplest DM interpretations of  recent
results from direct DM detection
experiments~\cite{Bernabei:2010mq,Aalseth:2011wp,Angloher:2011uu}.

Even more interestingly, the region of small mass splittings between squarks
and the LSP (where the annihilation fluxes are largest) is precisely the
region that is generically difficult to probe directly with collider searches
for squarks; this is because the energy stored in hadronic jets coming from
the decay $\tilde{q}\to\chi q$ becomes too small to pass the trigger selection
in this limit. In this sense, indirect DM searches are complementary to
collider searches and constrain a part of the parameter space that is not
directly accessible by either LEP, Tevatron or  LHC
(Fig.~\ref{fig:muconstraints}).\footnote{ Note that also in the situation
where the scalar tops are heavy, which generally helps to lift the lightest
neutral Higgs boson mass to the high value of $\sim125\,$GeV consistent with
recent indications found by both ATLAS and CMS, our results from indirect DM
searches would not change and constitute important independent limits on light
first or second generation squarks.} Depending on the cosmic-ray propagation
model, these limits can exclude Bino masses up to 45--80 GeV in the degenerate
case.  Ongoing AMS-02 measurements are expected to strengthen the limits on
the annihilation cross-section by up to an order of magnitude (see
\fex~Ref.~\cite{Evoli:2011id}), which will make it possible to probe Bino
masses of 100 GeV and beyond.

%%%%%%%%%%%%%%%%%%%%%%%%%%%%%%%%%%%%%%%%
%%%%%%%%%%%%%%%%%%%%%%%%%%%%%%%%%%%%%%%%
\medskip
\paragraph{Note added} During the final stages of this work, we became aware
of another project analyzing the annihilation of DM into $\bar q q g$ final
states~\cite{TAS}.

\paragraph{Acknowledgments}  We would like to thank Piero Ullio for providing
the antiproton propagation routines that were used in
\mbox{Ref.~\cite{Evoli:2011id},} which will eventually become publicly
available with the next release of \ds\ \cite{DS}.  We also would like to
thank Shoji Asai, Keisuke Fujii and Yasuhiro Shimizu for useful comments.
M.A.~and T.B.~acknowledge support from the German Research Foundation (DFG)
through Emmy Noether grant BR 3954/1-1.

\end{document}